\documentclass[twocolumn,final,natbib]{svjour3}          

\smartqed  

\usepackage{mathptmx}      
\usepackage{amsfonts,amssymb,amsmath,graphicx,amscd,ulem,times}
\usepackage{color}
\usepackage{array}
\usepackage[latin1]{inputenc}

\newcommand{\bi}{\begin{itemize}}

\newcommand{\ei}{\end{itemize}}
\newcommand{\be}{\begin{equation}}
\newcommand{\ee}{\end{equation}}
\newcommand{\bea}{\begin{eqnarray}}
\newcommand{\eea}{\end{eqnarray}}

\newcommand{\gdot}{\dot\gamma}

\newcommand{\phid}{\phi_\text{div}}

\newcommand{\Ca}{\textit{Ca}}

\newcommand{\mun}{$^{-1}$}

\DeclareTextSymbol{\degre}{OT1}{23}

\journalname{Rheologica Acta}

\begin{document}

\title{Mixtures of foam and paste: suspensions of bubbles in yield stress fluids
}

\titlerunning{Mixtures of foam and paste: suspensions of bubbles in yield stress fluids}

\author{Michael Kogan \and Lucie Ducloué \and Julie Goyon \and Xavier Chateau \and Olivier Pitois \and Guillaume Ovarlez}

\authorrunning{M. Kogan - L. Ducloué - J. Goyon - X. Chateau - O. Pitois - G. Ovarlez}

\institute{Université Paris-Est, Laboratoire Navier (UMR CNRS
8205), Champs-sur-Marne, France,
\email{guillaume.ovarlez@ifsttar.fr}}

\date{Received: date / Accepted: date}

\maketitle

\begin{abstract}
We study the rheological behavior of mixtures of foams and pastes,
which can be described as suspensions of bubbles in yield stress
fluids. Model systems are designed by mixing monodisperse aqueous
foams and concentrated emulsions. The elastic modulus of the
bubble suspensions is found to depend on the elastic capillary
number $\Ca_{_G}$, defined as the ratio of the paste elastic
modulus to the bubble capillary pressure. For values of $\Ca_{_G}$
larger than $\simeq0.5$, the dimensionless elastic modulus of the
aerated material decreases as the bubble volume fraction $\phi$
increases, suggesting that bubbles behave as soft elastic
inclusions. Consistently, this decrease is all the sharper as
$\Ca_{_G}$ is high, which accounts for the softening of the
bubbles as compared to the paste. By contrast, we find that the
yield stress of most studied materials is not modified by the
presence of bubbles. This suggests that their plastic behavior is
governed by the plastic capillary number $\Ca_{\tau_y}$, defined
as the ratio of the paste yield stress to the bubble capillary
pressure. At low $\Ca_{\tau_y}$ values, bubbles behave as
nondeformable inclusions, and we predict that the suspension
dimensionless yield stress should remain close to unity, in
agreement with our data up to $\Ca_{\tau_y}=0.2$. When preparing
systems with a larger target value of $\Ca_{\tau_y}$, we observe
bubble breakup during mixing, which means that they have been
deformed by shear. It then seems that a critical value
$\Ca_{\tau_y}\simeq0.2$ is never exceeded in the final material.
These observations might imply that, in bubble suspensions
prepared by mixing a foam and a paste, the suspension yield stress
is always close to that of the paste surrounding the bubbles.
Finally, at the highest $\phi$ investigated, the yield stress is
shown to increase abruptly with $\phi$: this is interpreted as a
`foamy yield stress fluid' regime, which takes place when the
paste mesoscopic constitutive elements (here, the oil droplets)
are strongly confined in the films between the bubbles.
\end{abstract}

\section{Introduction}\label{section_intro}

Many dense suspensions involved in industrial processes (concrete
casting, drilling muds, foodstuff transport...) and natural
phenomena (debris-flows, lava flows...) are yield stress fluids
\citep{Coussot2005}. These are usually very polydisperse systems
in which yield stress arises from the colloidal forces between the
smallest particles \citep{Wagner2012}, and is increased by the
presence of rigid noncolloidal particles
\citep{Mahaut2008a,Chateau2008}.

In addition, these materials often contain air bubbles.
This is the case, e.g., of crystal bearing
magmas \citep{Griffiths2000,Gonnermann2007}, or of aerated food
emulsions \citep{vanAken2001}. In industry, gas is typically
injected into materials to improve their performance or to make
them lighter. The fact that the latter materials exhibit a yield
stress is essential to prevent bubbles from rising
\citep{Dubash2004,Dubash2007,Sikorski2009}. Besides stability
issue, it is particularly important to understand the mechanical
behavior of these aerated materials. In many processes, gas
incorporation is obtained by mixing a foam with a paste. This is
the case, e.g., in plasterboard production, in which a foamed
plaster slurry is prepared by mixing an aqueous foam with a gypsum
plaster slurry, in foam concrete production by pre-foaming
methods, where aqueous foam is mixed with a base mix
\citep{Ramamurthy2009}, or in food processing when mixing beaten
egg whites and a batter to make aerated food products. It is thus
crucial to understand the impact of mixing a foam with a paste,
and more generally of adding bubbles to a yield stress fluid.

Even the simpler case of bubbles in viscous fluids has been the
subject of only a few experimental studies
\citep{Rust2002a,Rust2002b,Llewellin2002}. It has been shown that
the viscosity of suspensions of bubbles in a Newtonian fluid
depends on the applied shear stress $\tau$, and is characterized
by two different viscosity plateaus at low and high values of
$\tau$ \citep{Rust2002a}. At low applied stress, the viscosity of
the suspension is an increasing function of the bubble volume
fraction $\phi$; at high values of $\tau$, it decreases with
$\phi$. This phenomenon is attributed to interfacial phenomena and
to the ability of the bubbles to resist shear-induced deformation
\citep{Rust2002b}. At low applied stress, bubbles are not
deformed; they behave like rigid spherical particles with a slip
boundary condition, thus explaining the viscosity increase with
$\phi$. Above a critical stress, surface tension is no longer
sufficient to maintain the bubbles in a spherical shape; the
bubbles are deformed by shear, which leads to a decrease of the
viscosity with $\phi$. This transition from nondeformable to
deformable bubbles has indeed been shown to be driven by a
capillary number, which compares the shear stress and the
interfacial stress \citep{Rust2002a}. More details are given in
Sec.~\ref{subsection_theory_bubbles}.

Most studies of bubbles in yield stress fluids have focused on the
stability of these systems against coarsening and buoyancy, both
in specialized fields \citep{Dutta2004a,Dutta2004b,Ley2009} and
fundamental studies
\citep{Koczo1992,Turner1999,Dubash2004,Dubash2007,Sikorski2009,Goyon2010,Salonen2012}.
To our knowledge, the impact of bubbles on the rheological
properties of such systems has been studied only for specific
materials, but has not been the subject of fundamental studies.
\citet{Struble} have studied the effect of air entrainment on
cement pastes and concrete rheology. In these materials, an
air-entraining agent (a surfactant) is put in the paste; this
stabilizes the bubbles that are produced by agitating the paste.
Rheological measurements show an increase of the material yield
stress with the air bubble fraction, whereas its plastic viscosity
decreases; this is \textit{a priori} surprising because, according
to the results of \citet{Rust2002a}, the first observation would
imply that bubbles are not deformed by shear whereas the second
would imply that they are deformed. It is worth noting that the
bubbles obtained with air-entraining agents are likely to be very
polydisperse. This size dispersity is not controlled, which poses
at least two problems: (i) scale separation between the bubbles
and the cement particles is not ensured, which may not allow the
performance of classical micromechanical analyses by considering
the interstitial paste as a continuous medium, and (ii) in some
conditions, the largest bubbles may be deformable under shear
whereas the smallest are not, which can make the overall response
rather complex. Finally, it appears that the role of bubbles in
the rheology of yield stress fluids has yet to be fully
understood.

The case of rigid particles in yield stress fluids has been
investigated recently in several studies
\citep{AnceyJorrot2001,Geiker2002,Mahaut2008a,Mahaut2008b,Vu2010}.
\citet{Mahaut2008a} have performed studies on model suspensions of
monodisperse particles in various yield stress fluids
(concentrated emulsions, Carbopol gels, colloidal gels). They have
observed that their elastic modulus $G'(\phi)$ increases with the
particle volume fraction $\phi$, and is well fitted to a
Krieger-Dougherty equation, as classically observed for
suspensions of particles in linear materials. Their yield stress
$\tau_y(\phi)$ increases more moderately with $\phi$.
\citet{Mahaut2008a} have shown that the
dimensionless elastic modulus $G'(\phi)/G'(0)$ and the
dimensionless yield stress $\tau_y(\phi)/\tau_y(0)$ of these
systems are related through a simple relationship with no fitting
parameter, as predicted by \citet{Chateau2008}, thus leading to a
simple theoretical expression for $\tau_y(\phi)$ in agreement with
experimental data (more details are given in
Sec.~\ref{subsection_theory_rigid}). It is finally sufficient to
characterize only one rheological property of these materials to
predict the value of the other ones. These results, obtained on
model systems, have been shown to be applicable to more complex
systems such as model mortars made of rigid spherical particles in
cement pastes \citep{Mahaut2008b}.

A question that arises is how these results are changed when rigid
particles are replaced by gas bubbles, and what the role of bubble
deformability is on the rheological behavior of suspensions in
yield stress fluids. In particular, by analogy with suspensions of
bubbles in Newtonian fluids, we need to understand how the
deformability of bubbles is controlled when the material is
sheared. We also question the possible link between the linear and
nonlinear properties of these systems.

In this paper, we investigate the rheological behavior of
suspensions of bubbles in yield stress fluids. In
Sec.~\ref{section_theory}, we briefly review and discuss the
theoretical behavior of suspensions of rigid particles and bubbles
in linear (elastic or viscous) and nonlinear (plastic) materials.
We deal in particular with the issue of bubble deformability in
sheared yield stress fluids. In Sec.~\ref{section_materials}, we
present the model systems used in this work, made by mixing a
monodisperse foam and a model yield stress fluid, namely a
concentrated emulsion. The experimental results are presented in
Secs.~\ref{section_elastic} and~\ref{section_yield}. We first
study the change in the elastic properties of the yield stress
fluid in its solid regime as the foam fraction added to the
material (and thus bubble volume fraction) is progressively
increased (Sec.~\ref{section_elastic}). We then study the
evolution of the plastic properties (Sec.~\ref{section_yield}).
The study of viscoplastic properties is reserved for future work.

\section{Theory}\label{section_theory}

In the following, we consider monodisperse\footnote{Some aspects
of the linear and nonlinear behavior of polydisperse suspensions
are discussed in \citep{Vu2010}.} spherical inclusions of diameter
$d$ dispersed in materials at a volume fraction $\phi$ . When they
are dispersed in a linear material, namely a
Hookean material of elastic shear modulus $G'(0)$ or a Newtonian
material of viscosity $\eta(0)$, the property of interest is the
dimensionless linear response of the material $g(\phi)$ (i.e., the
dimensionless elastic modulus $G'(\phi)/G'(0)$ or the
dimensionless viscosity $\eta(\phi)/\eta(0)$). When they are
dispersed in a plastic material of yield stress $\tau_y(0)$, this
becomes the value of the dimensionless yield stress
$\tau_y(\phi)/\tau_y(0)$.

In this section, we first review the case of suspensions of solid
rigid inclusions (particles), which has already been thoroughly
investigated in the literature, and provides a solid framework to
better understand the behavior of deformable inclusions. We then
discuss the possible behavior of suspensions of gas inclusions
(bubbles), which is not yet well understood.

\subsection{Suspensions of rigid particles}\label{subsection_theory_rigid}

We first focus on suspensions of rigid particles. Particles are
considered as perfectly rigid, and a no-slip boundary condition is
assumed at the interface between the particles and the suspending
material.

\subsubsection*{Linear response}

In the dilute limit ($\phi<<1$), it is shown that
\citep{Larson1999} \bea g(\phi)=1+2.5\phi
\label{eq_linear_dilute_rigid} \eea Many theoretical expressions
exist for values of $\phi$ beyond the dilute limit
\citep{Stickel2005}. For isotropic suspensions, it is shown that
$g(\phi)$ should be higher than the Hashin-Shtrikman bound
\citep{Hashin1963} for any value of $\phi$: \bea
g(\phi)\geq\frac{1+\frac{3}{2}\phi}{1-\phi}\label{eq_linear_Hashin}\eea
Experimentally, the linear behavior of suspensions is usually
found to be consistent with the Krieger-Dougherty phenomenological
equation \citep{Stickel2005} \bea
g(\phi)=\frac{1}{(1-\phi/\phid)^{2.5\phid}}\label{eq_linear_Krieger_rigid}
\eea which complies with Eq.~\ref{eq_linear_dilute_rigid} in the
dilute limit. A value of $\phid=0.57$ was found experimentally for
isotropic suspensions \citep{Mahaut2008a}, and $\phid=0.605$ for
anisotropic suspensions (structured by a flow)
\citep{Ovarlez2006}.

\subsubsection*{Nonlinear response}

\citet{Chateau2008} have developed a simple micromechanical
approach to predict the nonlinear behavior of suspensions in yield
stress fluids. The starting point is that, if the particles do not
store any elastic energy (rigid limit), the local strain in the
interstitial material $\gamma_{\text{local}}(\phi)$ when a given
strain $\Gamma_{\text{macro}}$ is applied to the suspension can be
estimated at first approximation as \bea
\gamma_{\text{local}}(\phi)=\Gamma_{\text{macro}}\sqrt{g(\phi)\,/(1-\phi)}
\label{eq_gamma_local_rigid} \eea This information on the local
strain can be used to predict the value of the dimensionless yield
stress of the suspension, which finally implies that it is related
to its dimensionless elastic modulus through a simple relationship
with no fitting parameter \citep{Chateau2008}:
\bea\tau_y(\phi)/\tau_y(0)=\sqrt{(1-\phi)\,g(\phi)}\label{eq_g_vs_tau_rigid}
\eea In the dilute limit, the dimensionless yield stress is thus
obtained by combining Eqs.~\ref{eq_linear_dilute_rigid}
and~\ref{eq_g_vs_tau_rigid}:
\bea\tau_y(\phi)/\tau_y(0)=1+\frac{3}{4}\phi\label{eq_tau_rigid_dilute}
\eea Combining Eqs.~\ref{eq_linear_Krieger_rigid}
and~\ref{eq_g_vs_tau_rigid} finally yields a simple
phenomenological law
\bea\tau_y(\phi)/\tau_y(0)=\sqrt{(1-\phi)\,(1-\phi/\phid)^{-2.5\phid}}\label{eq_tau_rigid_krieger}
\eea which is in good agreement with the observations of
\citet{Mahaut2008a} for $\phi\leq50\%$, thus validating
Eqs.~\ref{eq_gamma_local_rigid} and~\ref{eq_g_vs_tau_rigid}.

Similarly, when the interstitial yield stress fluid behaves as a
Herschel-Bulkley material $\tau=\tau_y+\eta_{_{HB}}\gdot^n$, it is
shown that a simple relationship exists between the dimensionless
yield stress and the dimensionless consistency
$\eta_{_{HB}}(\phi)/\eta_{_{HB}}(0)$ \citep{Chateau2008}.

\subsection{Suspensions of bubbles}\label{subsection_theory_bubbles}

\subsubsection*{Linear response}

For suspensions of bubbles in linear materials, two limiting cases
are well known: that of nondeformable bubbles (with infinite
surface tension to bubble diameter ratio), and that of fully
deformable bubbles (with no surface tension). In all cases, a slip
boundary condition is assumed at the interface between the bubbles
and the interstitial material.

For nondeformable bubbles, the linear response in the dilute limit
is \citep{Larson1999} \bea g(\phi)=1+\phi
\label{eq_linear_nondeformable_dilute} \eea The difference with
rigid particles (Eq.~\ref{eq_linear_dilute_rigid}) comes from the
slip boundary condition at the bubble interface. For fully
deformable bubbles, in the dilute limit, $g(\phi)$ is \bea
g(\phi)=1-\frac{5}{3}\phi \label{eq_linear_deformable_dilute} \eea
Little is known about the behavior at higher $\phi$. Theoretical
bounds nevertheless exist. For isotropic suspensions of fully
deformable bubbles, $g(\phi)$ should be lower than the Mori-Tanaka
bound \citep{Dormieux2006} at any $\phi$ \bea
g(\phi)<\frac{1-\phi}{1+\frac{2}{3}\phi}\label{eq_linear_deformable_Mori}\eea
Overall, $g(\phi)$ is expected to increase with $\phi$ for
nondeformable bubbles and to decrease with $\phi$ for fully
deformable bubbles, consistent with the observations of
\citet{Rust2002a}. Between these two limiting cases, the
deformability of bubbles should depend on the resistance offered
by the bubbles to shear-induced deformation, due to the air/fluid
surface tension $\sigma_t$. The relevant dimensionless number
involved in the bubble deformability should then be a capillary
number $\Ca$, constructed by comparing the characteristic stress
$4\sigma_t/d$ due to surface tension to a characteristic applied
shear stress.

For bubbles suspended in a Newtonian fluid of viscosity
$\eta_{_0}$ sheared at a shear rate $\gdot$ , the only
characteristic shear stress is $\eta_{_0}\gdot$. It has indeed
been shown \citep{Rust2002a} that \bea
\Ca_\eta=\frac{\eta_{_0}\gdot}{2\sigma_t/d}\label{eq_capillary_eta}\eea
drives the transition between nondeformable bubbles (viscosity
increasing with $\phi$) and deformable bubbles (viscosity
decreasing with $\phi$). This transition occurs at
$\Ca_\eta\approx1$.

To our knowledge, no rigorously derived theoretical expression
valid for all values of $\Ca_\eta$ exists in the literature. The
full time-dependent tensorial behavior of suspensions of viscous
drops has been calculated by \citet{Frankel1970} in the dilute
limit, in the framework of a first-order perturbation analysis:
drops are assumed to be only slightly deformable. In the case of
suspensions of bubbles, this theory applies only at low $\Ca_\eta$
and is written as a first-order expansion in $\Ca_\eta$. For
steady-state flows in simple shear, the dimensionless viscosity
reduces to Eq.~\ref{eq_linear_nondeformable_dilute}: it does not
depend on $\Ca_\eta$ at order 1; in addition, normal stress
differences proportional to $\Ca_\eta$ are predicted: we will not
discuss this aspect in this paper. Several works
\citep{Rust2002a,Llewellin2002,Pal2004} have proposed
phenomenological equations intended to be valid at any values of
$\Ca_\eta$ and $\phi$. Their starting point is \citet{Frankel1970}
equation, with the additional strong assumption that this equation
can be used at any value of $\Ca_\eta$. Some of these expressions
have been shown to fit reasonably well experimental data; in
Sec.~\ref{subsection_elastic_fixed_capillary}, we will use the
best fitting function to experimental data proposed by
\citet{Rust2002a} and \citet{Pal2004} for comparison with our
data.

The case of bubbles with surface tension in elastic materials has
not been studied to our knowledge. Two capillary numbers can
\textit{a priori} be built to describe the behavior of such
materials, one based on the elastic modulus of the material
$G'(0)$: \bea
\Ca_{_G}=\frac{G'(0)}{2\sigma_t/d}\label{eq_capillary_g}\eea and
one based on the applied stress $\tau=G'(0)\Gamma$ (where $\Gamma$
is the applied strain) during an elastic modulus measurement: \bea
\Ca_\tau=\frac{G'(0)\Gamma}{2\sigma_t/d}\label{eq_capillary_g_gamma}\eea
For the materials we study (Sec.~\ref{subsection_materials}):
$2\sigma_t/d$ ranges between 25 and 1000~Pa, $G'$ is typically of
order 100 to 1000~Pa, and the applied stress during an elastic
modulus measurement is typically of order 1~Pa or less. The two
capillary numbers defined above would thus lead to descriptions of
very different behaviors. On the one hand, $\Ca_\tau$ is of order
0.01 for all studied systems, which would imply that bubbles
should behave as nondeformable objects; on the other hand,
$\Ca_{_G}$ is of order unity, which would imply that a transition
between the behavior of a suspension of nondeformable bubbles and
that of a suspension of deformable bubbles should be observed. In
our experiments, by characterizing the elastic moduli of
suspensions of various compositions, we thus expect to be able to
identify the relevant capillary number.

\subsubsection*{Nonlinear response}

Not much is known about the plastic behavior of suspensions of
bubbles in plastic materials. For nondeformable and fully
deformable bubbles, no elasticity is stored in the interfaces.
Eqs.~\ref{eq_gamma_local_rigid} and~\ref{eq_g_vs_tau_rigid} are
thus expected to remain valid. Combining
Eq.~\ref{eq_g_vs_tau_rigid} with
Eqs.~\ref{eq_linear_nondeformable_dilute}
and~\ref{eq_linear_deformable_dilute}, it is thus predicted that,
in the dilute limit, the dimensionless yield stress of suspensions
of nondeformable bubbles should be
\bea\tau_y(\phi)/\tau_y(0)=1\label{eq_plastic_nondeformable_dilute}\eea
whereas the dimensionless yield stress of suspensions of fully
deformable bubbles should be
\bea\tau_y(\phi)/\tau_y(0)=1-\frac{4}{3}\phi\label{eq_plastic_deformable_dilute}\eea
At this stage, as elasticity is stored in the interfaces for
intermediate cases, we cannot tell anything about the value of the
dimensionless yield stress and about its possible link with
$g(\phi)$ between these two limiting cases. \noindent In the case
of plastic flows, the characteristic stress is the interstitial
material yield stress $\tau_y$ and the capillary number that
should drive the transition from nondeformable to deformable
bubbles is expected to be \bea
\Ca_{\tau_y}=\frac{\tau_y}{2\sigma_t/d}\label{eq_capillary_tau_y}\eea

\section{Materials and methods}\label{section_materials}

\subsection{Materials}\label{subsection_materials}

\subsubsection*{Principle}

Model suspensions of bubbles in a yield stress fluid are prepared
by mixing a monodisperse foam and a concentrated emulsion. The
emulsion is made of oil droplets dispersed at a high volume
fraction (typically 82\%) in an aqueous surfactant solution; this
is a simple yield stress fluid \citep{Mason1996,Ovarlez2008}. The
foam is a dispersion of monodisperse bubbles at a high volume
fraction (typically 90\%) in an aqueous surfactant solution of the
same composition as that in the concentrated emulsion. The
similarity of the continuous phases of both systems ensures (i)
that the two materials are easily mixed together, and (ii) that
after mixing, the bubbles are surrounded by a known yield stress
fluid. This interstitial fluid is a concentrated emulsion of
droplet volume fraction slightly lower than that of the initial
emulsion, which has been diluted by the aqueous solution brought
in by the foam. All systems are such that (i) the droplet volume
fraction is high enough in the final interstitial emulsion so that
it has a yield stress, and that (ii) the bubble diameter is much
larger than the droplet diameter to ensure that the emulsion is
`seen' as a continuous material (a yield stress fluid) by the
bubbles.

\subsubsection*{Emulsion preparation}

Batches of 2~l of oil-in-water emulsion are prepared by dispersing
dodecane (Acros Organics) at 82\% or 85\% volume fraction in a
surfactant solution (see below) with a Silverson L4RT mixer. The
rotation speed is kept at 1000~rpm during the addition of
dodecane. It is then increased to 6000~rpm during $\approx30$~min
until a homogeneous emulsion is formed. The average droplet
diameter in the final emulsion is $4.2\mu$m; the polydispersity
(computed as in \citep{Mabille2000}) is of order 20\%. The
viscosity of dodecane is $\eta_o=1.3$~mPa.s at 25\degre\!C.

\subsubsection*{Foam}

Monodisperse foams are produced by either blowing air through a
porous glass frit or through a needle into a glass syringe filled
with a surfactant solution. The needle tip was cut and crushed:
depending on the way the tip was crushed, a different bubble size
was obtained for each needle; in the case of the porous frits,
different bubble sizes were obtained depending on the pore size.
In both cases, the air was blown at a flow rate of the order of a
few ml/min (depending on the bubble size), and was saturated with
perfluorohexane vapor (C$_6$F$_{14}$, Sigma-Aldrich), which
strongly reduces the bubble coarsening rate \citep{Gandolfo1997}.
We make monodisperse foams of bubble diameter varying between
110~$\mu$m and 3~mm (with a size dispersity of
order 3\%), which is much larger than the emulsion droplet
diameter. The size of the bubbles was checked by squeezing out a
small quantity of the foam into a petri dish filled with
surfactant solution and examining them under an upright
microscope. Slightly polydisperse foams of 100$\pm30\mu$m bubble
diameter were also produced by blowing air through a porous medium
made of a glass bead assembly: this setup allows us to produce a
large volume of foam made of small bubbles in a reasonable time,
that is, before coarsening occurs.

\subsubsection*{Surfactant solutions}

The surfactant solutions used in these studies are made of sodium
dodecyl sulfate surfactant (SDS, Sigma-Aldrich) dispersed in
deionized water at a 28~g/l concentration in the emulsions, and 5
g/l in the foams. The air/surfactant solution surface tension
$\sigma_t$ was measured with the pendant drop method; its value is
0.036$\pm$0.001~N.m\mun. The surfactant concentration has been
chosen to be high enough to saturate the oil droplet and bubble
interfaces. Note that the critical micelle
concentration (CMC) of the SDS solution is 2.3~g/l.

Complementary experiments have been performed with two other
surfactant solutions (both in the emulsion and the foam) in order
to study the role of surface mobility \citep{Denkov2009}:
TetradecylTrimethylAmmonium Bromide (TTAB) and a TTAB solution
with dodecanol. Those systems exhibited the same behavior as those
prepared with SDS; in this paper, we thus present only the results
obtained in systems prepared with the SDS surfactant.

\subsubsection*{Preparation of suspensions of bubbles}

\begin{figure}[btp] \begin{center}
\includegraphics[width=8.5cm]{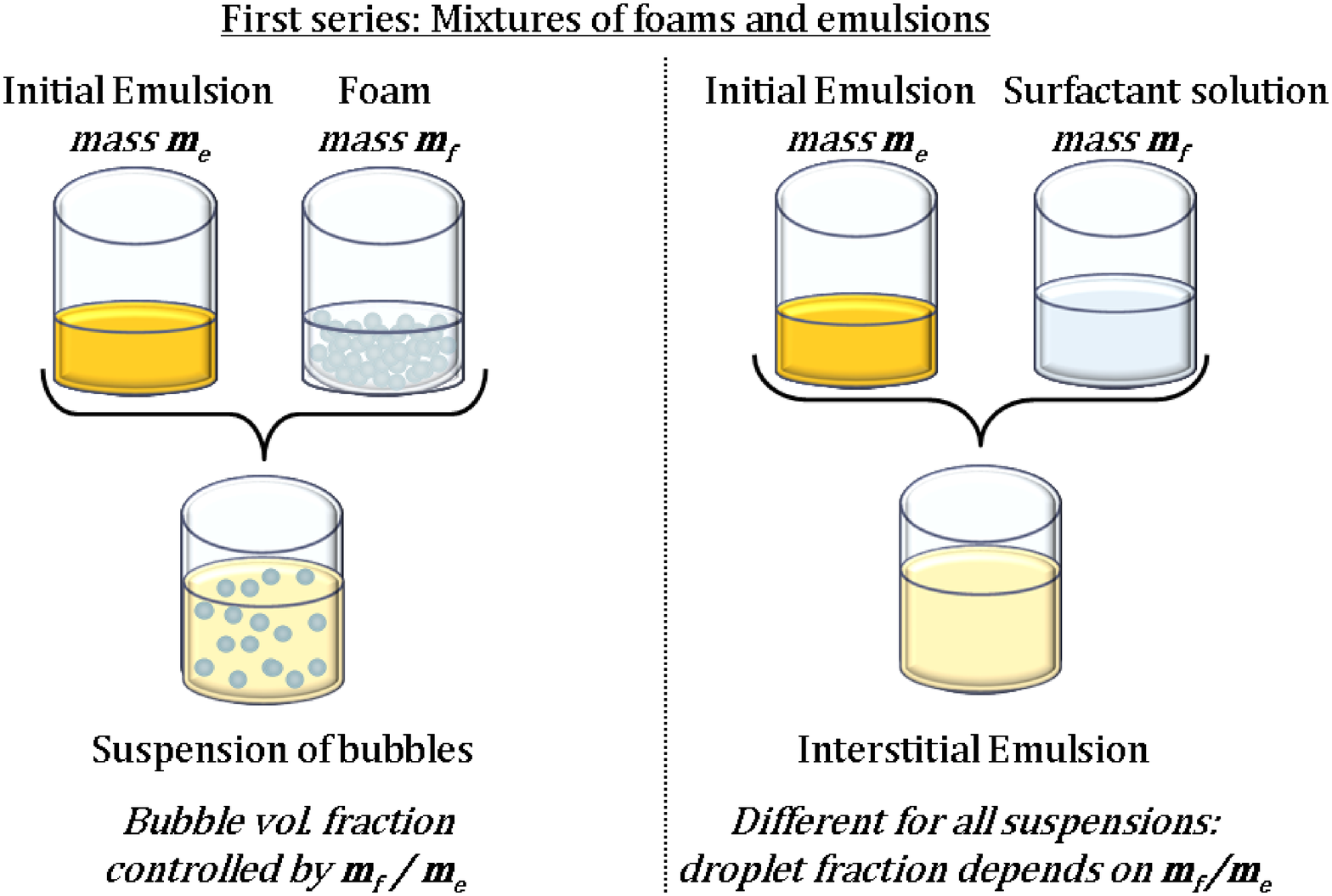}\\ \ \\
\includegraphics[width=8.5cm]{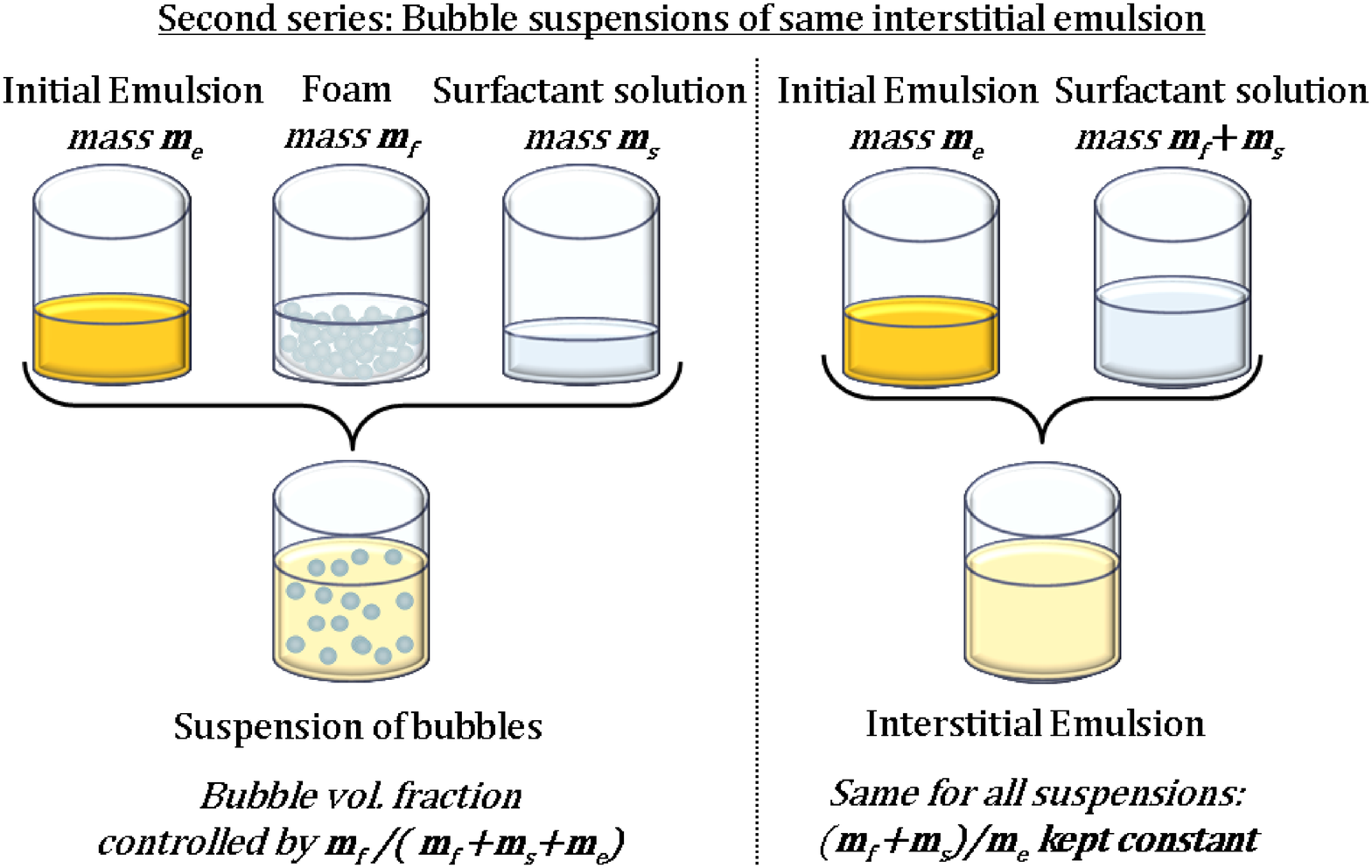}
\caption{Sketch of the material preparation in the two series of
experiments.}\label{fig_suspension_preparation}
\end{center} \end{figure}

Two kinds of studies are performed. First, we investigate the
evolution of the rheological behavior of a yield stress fluid when
foam is added to the material (this matches what may happen in an
industrial process). In a first series of experiments, samples are
thus prepared by mixing a foam and an emulsion in various mass
ratios (Fig.~\ref{fig_suspension_preparation}a). In this case, as
the foam fraction is progressively increased, two changes occur in
the final material: (i) the bubble volume fraction increases, and
(ii) the droplet volume fraction in the interstitial emulsion
decreases (due to the additional surfactant solution brought in by
the foam). To understand the impact of adding bubbles to a yield
stress fluid, the properties of the bubble suspension have to be
compared to that of the yield stress fluid surrounding the
bubbles. For each sample, we thus also prepare a pure emulsion
sample by adding to the initial emulsion a mass of surfactant
solution equal to the mass of the foam added to make the
suspension sample. This allows us to characterize the interstitial
emulsion for each prepared suspension. In this series of
experiments, the elastic modulus of the initial emulsion (before
mixing with the foam) is $G'=625$~Pa and its yield stress is
$\tau_y=35$Pa. The bubble volume fraction $\phi$ is varied between
0\% and 90\%.

In a second series of experiments, we focus on the role of bubbles
by studying the impact of increasing the bubble volume fraction in
an interstitial yield stress fluid of constant properties
(Fig.~\ref{fig_suspension_preparation}b). To this end, materials
are prepared by adding various mixes of foam (mass $m_f$) and
surfactant solution (mass $m_s$) to the emulsion (mass $m_e$). The
ratio $\frac{m_f+m_s}{m_e}$ is kept constant, which ensures that
the final droplet volume fraction in the emulsion surrounding the
bubbles is always the same. The bubble volume fraction in the
material is then changed by varying $m_f/m_s$. All suspensions are
compared to the interstitial emulsion, which is made with the same
recipe, with $m_f/m_s=0$. In this series of experiments, the
elastic modulus of the interstitial emulsion is $G'\approx280$~Pa
and its yield stress is $\tau_y\approx9$~Pa. A few experiments
with the largest bubbles are also performed with interstitial
emulsions of yield stress equal to 19 and 72~Pa to better test the
possible deformability of bubbles under shear. The bubble volume
fraction $\phi$ is here varied between 0\% and 60\%.

The foam, surfactant solution, and the emulsion are mixed together
in a 9.5~cm diameter beaker using a mixer with a 6.8cm x 6.8cm,
6-hole rectangular blade. The velocity is initially at 60 rpm, and
is increased to 100~rpm after $\approx$10s of mixing. The mixing
then continues until the mixture is homogeneous, which occurs
after $\approx$60~s.

The volume fraction of bubbles within the emulsion is estimated
from the measurement of the density of the suspension in a petri
dish of known volume. The absolute uncertainty
on the material volume fraction, accounting for both the
measurement uncertainty and the possible volume fraction spatial
variations in a given sample, is 0.02. We disregarded the few
samples where the experimentally measured bubble fraction was
greater than that expected from the foam fraction in the mixture,
to ensure that no unwanted bubbles were added to the material.

All systems are designed to be stable at rest: their yield number
$Y=\frac{\tau_y}{\Delta\rho gd}$ complies with the stability
criterion of a single bubble embedded in a yield-stress fluid
\citep{Dubash2007}, which should ensure that elastic forces
exerted by the yield stress fluid at rest are able to
counterbalance the net gravity force.

We checked on several prepared suspensions that (i) this procedure
leads to a homogeneous material, by measuring the density of
several samples extracted from different heights in a given
material, and that (ii) the bubble size in the suspension is equal
to that in the initial foam, i.e., there is neither shear-induced
coalescence nor breakup during mixing, except in the cases
discussed at the end of Sec.~\ref{section_yield}, and no
significant bubble coarsening occurs at the time scale of the
experiments.

\subsection{Rheometry}\label{subsection_rheometry}

Rheometric experiments are performed within a vane-in-cup geometry
(dimensions for bubbles of diameter $d\leq320~\mu$m: inner radius
$R_i=12.5$~mm, outer cylinder radius $R_e =18$~mm, height
$H=46$~mm; dimensions for $d=800~\mu$m to $d=3$~mm: $R_i=22.5$~mm,
$R_e=45$~mm, $H=46$~mm) on a commercial rheometer (Bohlin C-VOR
200) that imposes either the torque or the rotational velocity
(with a torque feedback). In order to avoid wall slip
\citep{Coussot2005}, we use a six-blade vane as an inner tool, and
we glue sandpaper on the outer cylinder wall. Working within these
wide-gap geometries allows for study of samples with large bubbles
and ensures that, for all the materials studied, there are enough
bubbles in the gap to consider that we are measuring the
properties of a continuous medium (the suspension). In such a
geometry, the shear stress distribution in the gap is
heterogeneous. Therefore, one has to choose a definition of the
shear stress $\tau$ that is measured in a given rheological
experiment. Here, we want to perform both elastic modulus and
yield stress measurements. Whatever the measurement method we
choose, yield first occurs where the stress is maximal i.e. along
the inner virtual cylinder delimited by the blades\footnote{Note
that although flow does not generally have an azimuthal symmetry
in a vane-in-cup geometry \citep{Baravian2002,Ovarlez2011},
azimuthal symmetry seems to be recovered when the shear stress is
close to the yield stress \citep{keentok1985,Ovarlez2011})}. As a
consequence, the contribution to the torque $T$ of the sheared
material at yield is: $\tau_y*2\pi HR_i^2$; as the material is
also sheared at the bottom of the vane, another contribution to
the torque is given by $\tau_y*2\pi R_i^3/3$ \citep{nguyen1983}.
We thus define the shear stress measurement as
$\tau(R_i)=\frac{T}{2\pi HR_i^2(1+R_i/3H)}$, so that the yield
stress $\tau_y$ is correctly measured; this is validated by
comparison with measurements performed in a cone-and-plate
geometry. To ensure correct elastic modulus measurement, as the
strain field does not have azimuthal symmetry in the elastic
regime, we calibrate the conversion factor between the rotation
angle $\Theta$ and the strain $\gamma$ to ensure that the same
elastic modulus is measured in a cone-and-plate geometry and in
the vane-in-cup geometry; due to the linearity of the behavior,
this calibration made in a particular linear elastic material
remains valid for any linear elastic material.

\subsection{Procedure, elastic modulus and yield stress measurements}

Before designing the experimental procedure, we need to define
precisely the state of the materials we want study. Two points are
important: (i) measurements should be performed on a homogeneous
suspension, and (ii) the microstructure of the suspensions should
be controlled. These points impose severe restrictions on the
preparation and the yield stress measurement procedure, as
measurements involving an significant flow of suspensions (a large
strain) pose several problems.

First, flow causes particle migration
\citep{Leighton1987,Phillips1992,Ovarlez2006} towards the low
shear zones (the outer cylinder in coaxial cylinder geometries)
i.e. creation of a heterogeneous structure; in addition, with a
vane tool, particle depletion is induced near the vane blades
\citep{Ovarlez2011} leading to the equivalent of a slip layer. The
same might happen with bubbles. In density-mismatched suspensions,
even those stable at rest, another source of heterogeneity is
gravity \citep{Ovarlez2010,Ovarlez2012}: bubbles tend to rise when
the material is sheared, and rise faster and faster as the shear
rate is increased \citep{Goyon2010}. To circumvent these problems,
we have decided (i) to avoid any preshear of the material, and
(ii) to measure the static yield stress; any other yield stress
measurement method based on a shear flow such as shear rate
\citep{Geiker2002b} or shear stress ramps \citep{uhlherr2005} and
creep tests \citep{Coussot2006} may lead to heterogeneities. We
have also decided to postpone the study of viscous dissipation in
these materials, which would require special care and may require
the use of MRI methods to measure local volume fractions
\citep{Goyon2010}. Therefore, we here focus on the impact of
bubbles on the elastic modulus and the static yield stress of
yield stress fluids.

Another problem is that an anisotropic microstructure
(distribution of neighbors) is created when suspensions of
particles flow \citep{Gadala1980,parsi1987}; in addition, with
deformable bubbles, bubble orientation is expected to depend on
shear history \citep{Rust2002b}. Suspensions of isotropic and
anisotropic microstructure have very different rheological
properties \citep{Blanc2011}; one thus has to be careful about
shear history when characterizing a suspension, to ensure that the
same structure is always dealt with. Here we choose to
characterize the material as prepared, which we expect to be
roughly isotropic given the complexity of the mixing flow. This
point is also in favor of the use of a vane-in-cup geometry to
study the materials: the use of a vane allows the study of the
properties of the prepared material with minimal disturbance of
the material structure during the insertion of the tool
\citep{nguyen1983}.

\noindent Finally, following \citet{Mahaut2008a}, the procedure is
the following:
\begin{itemize}
\item a suspension is prepared as detailed is
Sec.~\ref{subsection_materials}, and poured into the cup of the
vane-and-cup geometry. The vane tool is then slowly inserted into
the material.

\item the elastic modulus G' is determined through oscillatory
shear experiments. As we work with a controlled stress rheometer,
a shear stress is imposed rather than a shear strain, in order to
get accurately small deformations. The oscillatory stress is
imposed during 1~min at a 2~Hz frequency. Its amplitude
$\tau_{_0}$ depends on the sample and is chosen so as to ensure
that the strain induced on the tested material is lower than
$10^{-3}$ and that all materials are tested in their linear
regime. The independence of the results on the choice of
$\tau_{_0}$ was checked on some materials. We checked the
independence of the results on the frequency in the 0.1-10 Hz
range: although the elastic modulus $G'(0)$ of the emulsion
depends on the frequency, the dimensionless modulus
$G'(\phi)/G'(0)$ does not significantly depend on it; the impact
of bubbles is thus properly accounted for. A more detailed study
of a possible frequency dependence (which can be expected due to
the presence of interfaces) is postponed to another study.

\item afterwards, we perform our yield stress measurement with the
vane method \citep{nguyen1985,Liddell1996}: a small rotational
velocity, corresponding to a 0.01~s\mun macroscopic shear rate, is
imposed on the vane tool during 120~s. We checked that the same
effect of the bubbles on the yield stress is observed whatever the
low velocity that is chosen to drive the vane tool. We also
checked that the elasticity measurement is nonperturbative: the
same yield stress is measured if a zero stress is imposed instead
of oscillations before the yield stress measurement

\item finally, as the yield stress measurement may induce
migration, bubble rising and microstructure anisotropy, any new
measurement requires new sample preparation.
\end{itemize}

\begin{figure}[btp] \begin{center}
(a)\includegraphics[width=6cm]{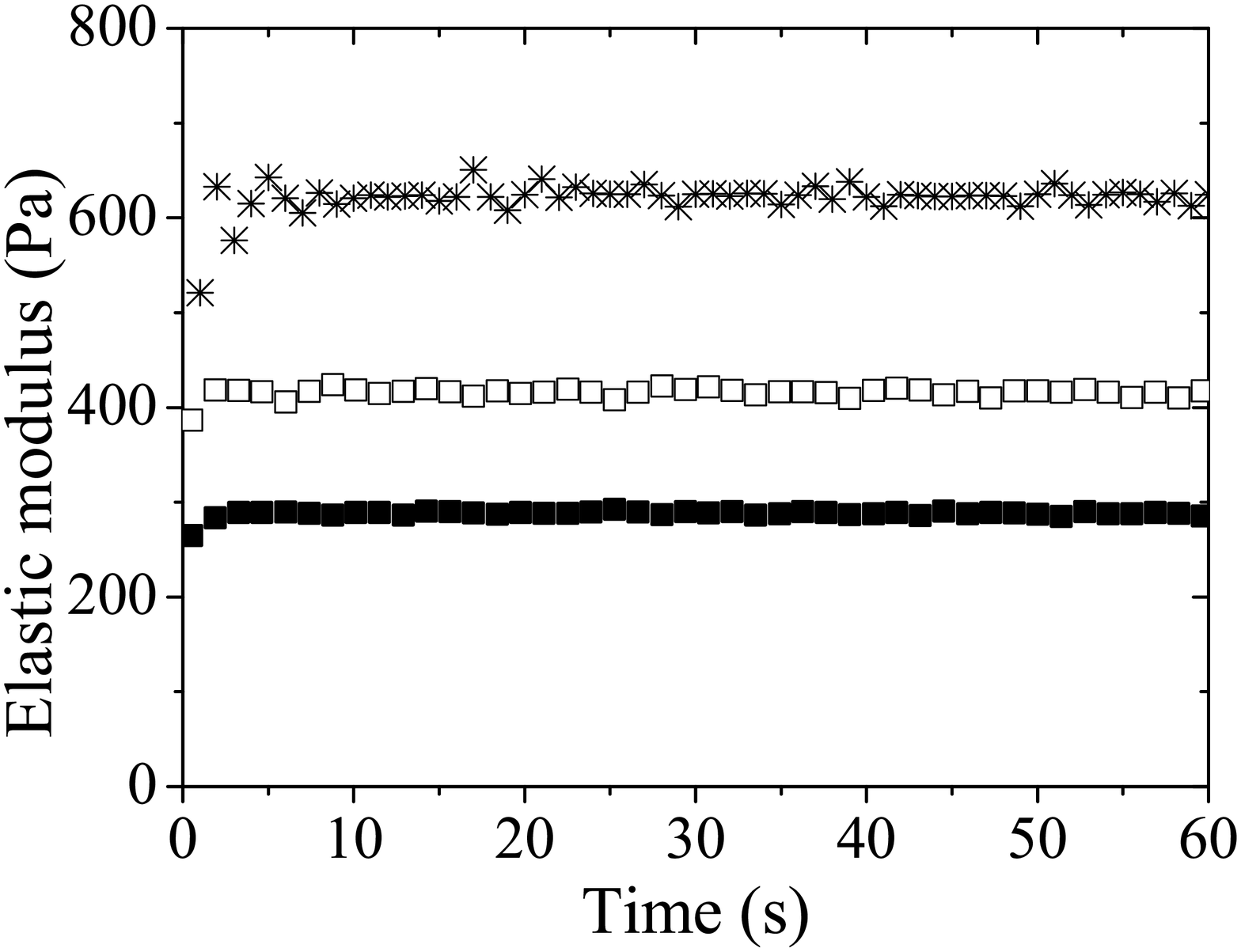}\hspace{0.4cm}\, \\
(b)\includegraphics[width=6cm]{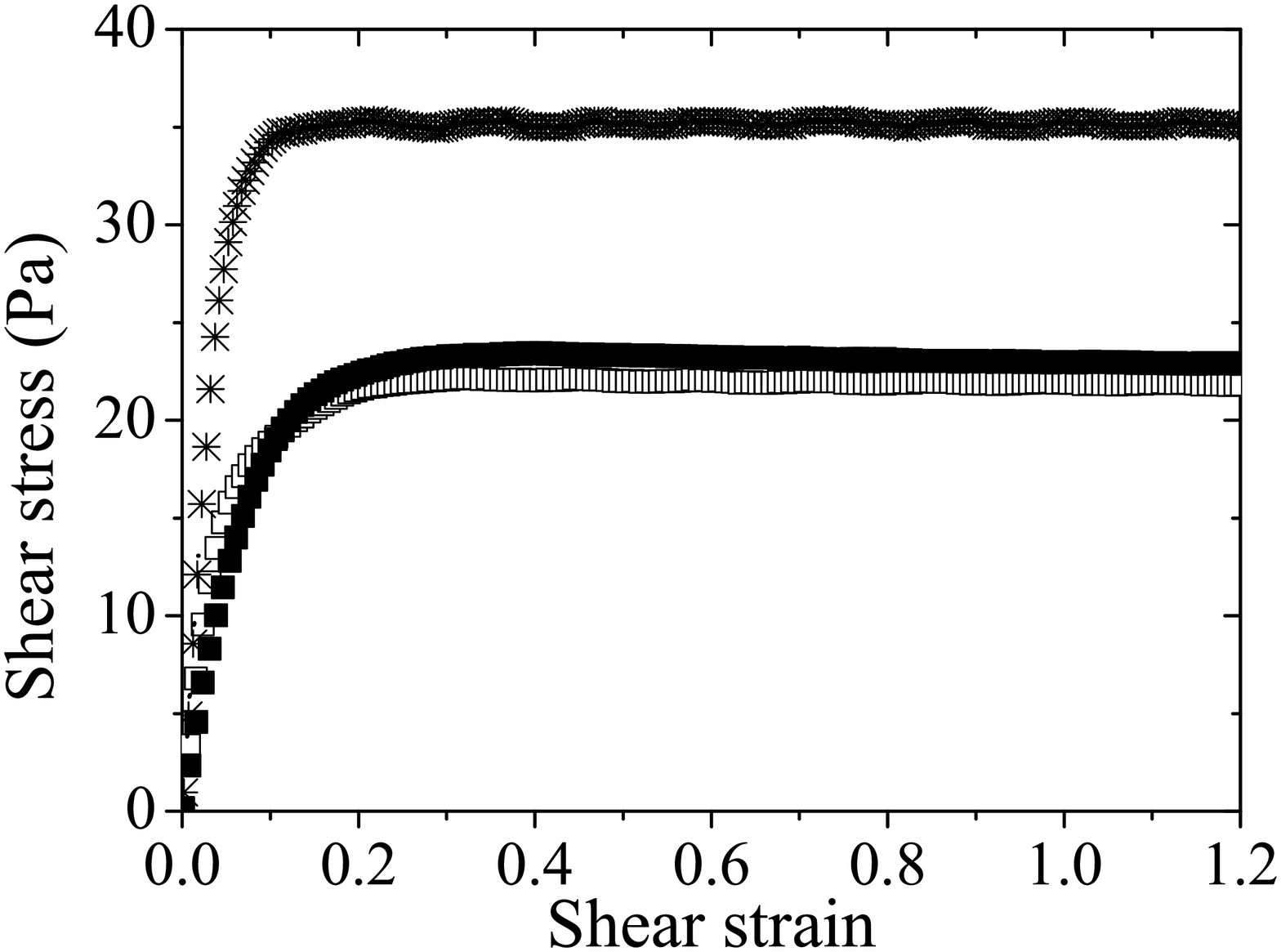}\hspace{0.4cm}\, \\
\caption{(a) Elastic modulus G' vs. time after loading when an
oscillatory shear stress of amplitude $\tau_{_0}=0.35$~Pa is
applied at a frequency of 2~Hz to a pure emulsion (crosses), a
suspension of bubbles in the emulsion (filled squares), and in the
interstitial emulsion of the suspension (diluted emulsion, empty
squares). The bubble volume fraction in the suspension is 43\%,
the bubble size is 320~$\mu$m. (b) Shear stress vs. shear strain
when slowly shearing the same materials from rest at $10^{-2}
s^{-1}$.}\label{fig_procedure_gprime_seuil}
\end{center} \end{figure}

Fig.~\ref{fig_procedure_gprime_seuil} shows raw elastic and yield
stress measurements performed in a pure emulsion, in a suspension
of 320~$\mu$m bubbles (made by adding foam to this last emulsion)
and in its corresponding interstitial emulsion (made by adding
only surfactant solution to the emulsion). $G'$ remains constant
in time, consistent with the nonthixotropic character of
emulsions. This indicates that the only impact of adding
surfactant solution and bubbles to the emulsion is a change in the
$G'$ value: no new significant mechanism of aging has arisen. The
elastic modulus $G'$ of the material is thus unambiguously defined
from this experiment; we study its dependence on the material
composition in Sec.~\ref{section_elastic}.

On the stress vs. strain plot during the yield stress measurement
(Fig.~\ref{fig_procedure_gprime_seuil}b), there is first a linear
increase of stress with strain: this corresponds to the elastic
deformation of the material. There is then a well-defined plateau,
corresponding to the plastic flow at low shear rate
(we checked that viscous effects are here
negligible), as usually observed in simple yield stress fluids.
This plateau defines unambiguously the yield stress $\tau_y$ of
the materials; its dependence on the material composition is
studied in Sec.~\ref{section_yield}.

As observed in
Fig.\ref{fig_procedure_gprime_seuil}, the uncertainty on the
measurement of the elastic modulus and of the yield stress of a
given sample is very low (less than 1\%). The uncertainty on the
suspension mechanical properties mostly comes from the
reproducibility of the whole sample preparation procedure (and
most probably from the dilution of the interstitial emulsion).
Since the liquid fraction in the aqueous foam is not controlled,
our procedure does not allow us to obtain several samples at the
same exact bubble volume fraction. We are thus not able to
evaluate rigorously this uncertainty. Nevertheless, from several
couples of data obtained in materials prepared at approximately
the same volume fraction, we can evaluate the uncertainty on the
suspension elastic modulus and yield stress to be of order 5\%.

\section{Elastic modulus}\label{section_elastic}

\subsection{Mixtures of foam and emulsion}\label{subsection_elastic_mixes}

In this section, we first study the cases where suspensions of
increasing bubble volume fraction are obtained by adding foam to a
given emulsion in increasing foam to emulsion mass ratios. We
recall that each suspension is compared to the emulsion
surrounding the bubbles, made by adding only surfactant solution
to the initial emulsion (see Sec.~\ref{subsection_materials},
Fig.~\ref{fig_suspension_preparation}).

\begin{figure*}[btp]\begin{center}
\hspace{0.5cm}(a)\hspace{-0.5cm}\includegraphics[width=12.8cm]{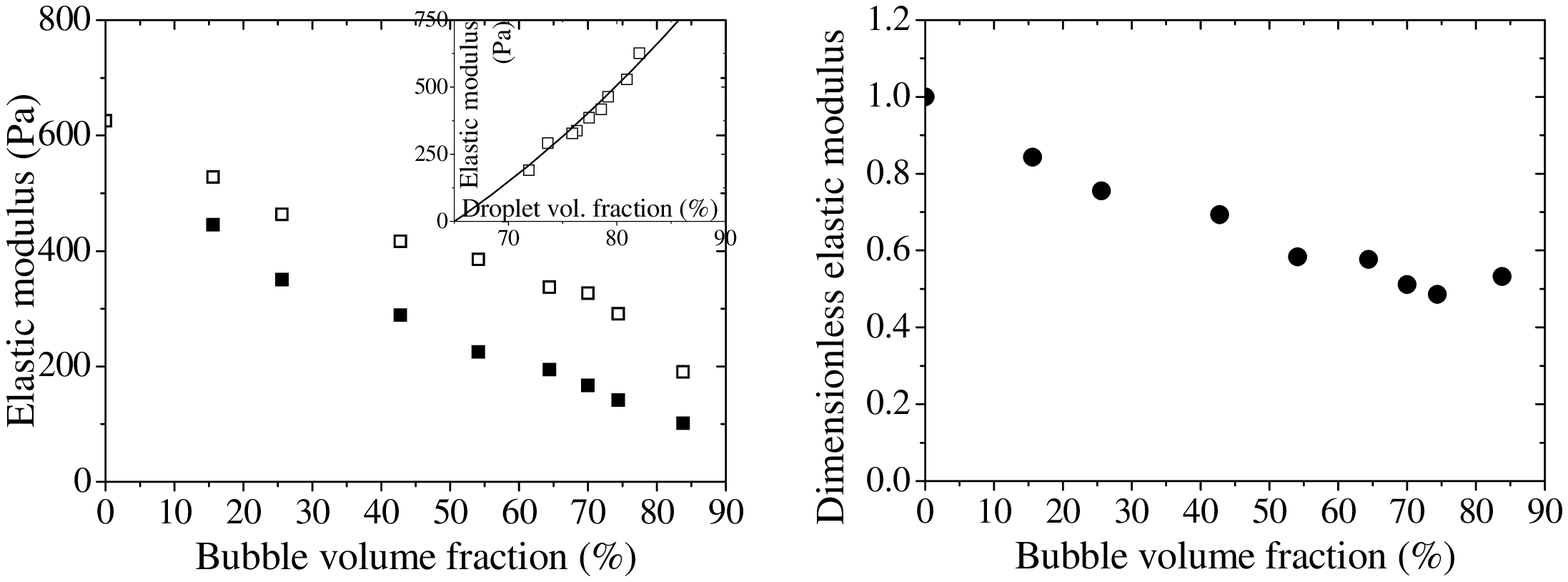}\hspace{-0.3cm}(b)\hspace{0.3cm}\, \\
\caption{(a) Elastic modulus vs. bubble volume fraction for the
suspensions of bubbles in the emulsion (filled squares), and for
the emulsions surrounding the bubbles in the suspensions (diluted
emulsions, empty squares). (b) Dimensionless elastic modulus
$g(\phi)$ vs. bubble volume fraction $\phi$. The bubble size is
320~$\mu$m. It is recalled that an increasing $\phi$ corresponds
to an increasing quantity of foam added to the emulsion, and to a
lower droplet volume fraction in the interstitial emulsion. Inset:
elastic modulus $G'_i$ of the interstitial emulsion vs. oil
droplet volume fraction $\phi_d$ in the emulsion. The line is a
fit to the $\phi_d(\phi_d-\phi_m)$ scaling proposed by
\citet{Mason1995} with
$\phi_m=65\%$.}\label{fig_gprime_phi_bulles320}
\end{center}\end{figure*}

In Fig.~\ref{fig_gprime_phi_bulles320}a we plot the values of the
elastic modulus $G'$ measured in suspensions of 320~$\mu$m bubbles
and in their corresponding interstitial emulsion, as a function of
the bubble volume fraction $\phi$ in the suspension. As expected
from the dilution of the material by the surfactant solution, the
more foam we incorporate (i.e., as $\phi$ increases), the lower
the elastic modulus of the interstitial fluid gets. In
concentrated emulsions, $G'$ is actually known to be a monotonic
increasing function of the droplet volume fraction above the
jamming packing fraction $\phi_m$ \citep{Mason1995} and to tend to
zero at $\phi_m$. The inset of
Fig.~\ref{fig_gprime_phi_bulles320}a shows the elastic modulus of
the interstitial emulsion replotted as a function of the droplet
volume fraction $\phi_d$ in the emulsion (higher quantities of
foam added to the system, and thus higher bubble volume fraction
$\phi$ in the suspension, corresponds to lower values of
$\phi_d$); the observed behavior is consistent with the law
proposed by \citet{Mason1995}.

The elastic moduli of the bubble suspensions are all significantly
lower than those of the interstitial emulsions. This means that
the impact of adding foam is twofold: (i) there is a decrease of
the elastic modulus of the interstitial fluid due to its dilution,
and (ii) there is an additional decrease of the modulus of the
suspension due to the addition of bubbles. To quantify the
decrease due solely to the presence of bubbles, we plot the
dimensionless elastic modulus $g(\phi)=G'(\phi)/G'_i$ vs. bubble
volume fraction $\phi$ in Fig.~\ref{fig_gprime_phi_bulles320}b;
here, $G'_i$ is the elastic modulus of the interstitial emulsion
corresponding to each suspension, i.e., accounting for dilution
effects. In this system (suspension of 320~$\mu$m bubbles),
$g(\phi)$ is found to decrease regularly (basically linearly) when
$\phi$ increases. E.g., for 75\% of bubbles, $G'$ is decreased by
a factor 2 due to the presence of bubbles.

\begin{figure}[btp]\begin{center}
\includegraphics[width=7.5cm]{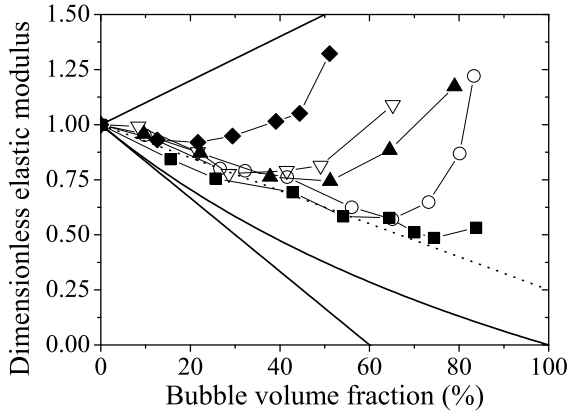}
\caption{Dimensionless elastic modulus $g(\phi)$ vs. bubble volume
fraction $\phi$, for suspensions of bubble diameter $d=$~320$\mu$m
(squares), 260$\mu$m (empty circles), 230$\mu$m (up triangles),
210$\mu$m (empty down triangles), and 110$\mu$m (diamonds). The
dotted line is a $g(\phi)=1-0.75*\phi$ equation. The full lines
correspond to, from bottom to top: the theoretical value of
$g(\phi)$ for suspensions of fully deformable bubbles in the
dilute limit (Eq.~\ref{eq_linear_deformable_dilute}), the
theoretical upper bound for isotropic suspensions of fully
deformable bubbles (Eq.~\ref{eq_linear_deformable_Mori}), and the
theoretical value of $g(\phi)$ for suspensions of nondeformable
bubbles in the dilute limit
(Eq.~\ref{eq_linear_nondeformable_dilute}).}\label{fig_gprime_ttesbulles_emulvariable}
\end{center}\end{figure}

In Fig.~\ref{fig_gprime_ttesbulles_emulvariable}, we now plot all
the dimensionless moduli $g(\phi)$ determined in the suspensions
of bubbles of diameter $d$ varying between 110 and 320~$\mu$m. Two
behaviors are observed. (i) For a volume fraction $\phi$ lower
than a critical value $\phi_c(d)$ that depends on $d$, $g(\phi)$
decreases with increasing $\phi$, and, at first glance, all the
data seem to collapse onto a roughly linear curve. (ii) Above
$\phi_c(d)$, $g(\phi)$ is found to increase with $\phi$, and the
way it increases depends strongly on $d$. At the highest volume
fraction reached in all systems, the elastic modulus of the bubble
suspension is up to 35\% higher than that of the interstitial
emulsion.

\subsection{Role of bubble deformability}\label{subsection_elastic_interpretation}

Let us now try to understand this behavior. As explained in
Sec~\ref{section_theory}, a decrease of the elastic modulus when
the bubble volume fraction is increased is the signature of bubble
deformability under shear. The fact that $G'$ changes from
decreasing to increasing with $\phi$ around a critical value
$\phi_c(d)$ would then mean that bubbles are changed from
deformable to nondeformable when the bubble volume fraction is
increased. This is \textit{a priori} surprising as the ability of
the bubbles to be deformed under shear should not depend on $\phi$
(as long as the material is not in a `foam' regime). However, it
should be recalled that, with the procedure used here to prepare
the material, increasing values of $\phi$ imply decreasing values
of the interstitial material elastic modulus $G'_i$
(Fig.~\ref{fig_gprime_phi_bulles320}a). This suggests that the
transition is driven by the elastic capillary number
$\Ca_{_G}=G'_i/(2\sigma_t/d)$, which compares the material and
bubble stiffnesses (Sec.~\ref{subsection_theory_bubbles}).

\begin{figure}[btp]\begin{center}
\includegraphics[width=7cm]{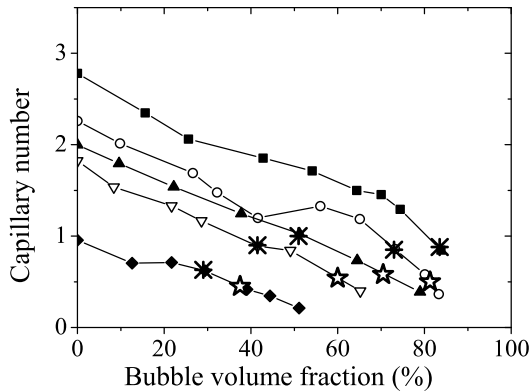}\\
\caption{Elastic capillary number $\Ca_{_G}$ vs. bubble volume
fraction $\phi$, for suspensions of bubble diameter $d=$~320$\mu$m
(squares), 260$\mu$m (empty circles), 230$\mu$m (up triangles),
210$\mu$m (empty down triangles), and 110$\mu$m (diamonds), in the
experiments of Fig.~\ref{fig_gprime_ttesbulles_emulvariable}. The
empty stars are $\Ca_{_G}(\phi)$ values corresponding to
$g(\phi)=1$ (interpolated from
Fig.~\ref{fig_gprime_ttesbulles_emulvariable}), i.e., to the
transition from soft to stiff bubbles. The crosses are
$\Ca_{_G}(\phi)$ values below which $g(\phi)$ data do not collapse anymore onto a single curve in
Fig.~\ref{fig_gprime_ttesbulles_emulvariable}.}\label{fig_capillary}
\end{center}\end{figure}

To go one step further, we thus plot the $\Ca_{_G}$ values vs. the
bubble volume fraction $\phi$ for all the studied systems
(Fig.~\ref{fig_capillary}); it is assumed here that the surface
tension $\sigma_t$ between the gas bubbles and the emulsion takes
the value of the air/surfactant solution surface tension. At low
$\phi$, we observe that $\Ca_{_G}$ is of order of a few units,
consistent with the decrease of $G'(\phi)$ with $\phi$: the
bubbles behave as soft inclusions (they are deformable). Below a
critical value of $\Ca_{_G}$ that is expected to be of order
unity, bubbles should progressively behave as rigid inclusions;
from Fig.~\ref{fig_capillary}, this should happen above a volume
fraction $\phi'_c(d)$ that depends on $d$, which may explain a the
$d$-dependent rise of $G'(\phi)$ at high $\phi$. To show that this
mechanism may account for our results, we plot in
Fig.~\ref{fig_capillary} the $\Ca_{_G}(\phi)$ values corresponding
to $g(\phi)=1$, i.e., to the transition from soft to stiff
bubbles; these values are interpolated from
Fig.~\ref{fig_gprime_ttesbulles_emulvariable}. A constant critical
capillary number $\simeq0.5$ seems to characterize this
transition; we will come back onto this point below. In
Fig.~\ref{fig_capillary}, we also plot the $\Ca_{_G}(\phi)$ values below which $g(\phi)$ data do not collapse anymore onto a single curve in
Fig.~\ref{fig_gprime_ttesbulles_emulvariable}. Again, a roughly
constant capillary number seems to characterize a change in the
elastic behavior; here, it likely marks the transition to a regime
where the behavior strongly depends on $\Ca_{_G}$. Note that this
approach seems to work at bubble volume fraction $\phi$ as high as
80\%, although the elastic capillary number is not expected to
play exactly the same role in the `foam' regime, for $\phi$ higher
than $\simeq64$\%, as in the `suspension' regime. Things can
indeed be more complex in a `foamy yield stress fluid', as bubbles
have to be deformed for purely geometrical reasons.

In the regime where bubble deformability is likely to be
important, the dimensionless elastic modulus $g(\phi)$ does not
follow the theoretical law for suspensions of fully deformable
bubbles in the dilute limit
(Eq.~\ref{eq_linear_deformable_dilute}): a $-5/3$ slope is indeed
expected at low $\phi$ whereas a slope of order of $-3/4$ is
observed (Fig.~\ref{fig_gprime_ttesbulles_emulvariable}). Moreover
the data fall above the theoretical upper bound of Mori-Tanaka
(Eq.~\ref{eq_linear_deformable_Mori}), expected to be valid for
bubbles with no surface tension. This indicates that we are not in
the limit where bubbles are fully deformable and that elasticity
must be stored in the interfaces due to surface tension. This
naturally leads to an increased rigidity of the system as compared
to the case with no surface tension; a theoretical upper bound for
this case still has to be computed. This observation is consistent
with the fact that the elastic capillary number $\Ca_{_G}$ is only
of the order of a few units in all the studied systems
(Fig.~\ref{fig_capillary}a).

At the lowest $\Ca_{_G}$ values, although the bubbles are expected
to start behaving as stiff inclusions, the dimensionless elastic
modulus $g(\phi)$ is well below the theoretical law for
suspensions of nondeformable bubbles in the dilute limit
(Eq.~\ref{eq_linear_nondeformable_dilute}). This would mean that
the bubbles are far from their nondeformability limit, consistent
with the fact that $\Ca_{_G}$ is only slightly less than unity in
these cases.

\subsection{Suspensions of bubbles at fixed capillary number}\label{subsection_elastic_fixed_capillary}

To better show that the elastic behavior is controlled by
$\Ca_{_G}$ and to quantify the impact of a change in $\Ca_{_G}$ on
the value of $g(\phi)$, we now study suspensions of bubbles
prepared with the same interstitial emulsion (of elastic modulus
$G'_{_0}=$285~Pa) between the bubbles at any $d$ and $\phi$ (see
Sec.~\ref{section_materials},
Fig.~\ref{fig_suspension_preparation}b). For a given bubble size,
the $g(\phi)$ curve is then characteristic of a given value of
$\Ca_{_G}$. We study suspensions made of bubbles of three
different diameters: 100$\mu$m, 300$\mu$m, and 1.6 mm; they are
characterized by capillary numbers $\Ca_{_G}$ equals respectively
to 0.4, 1.2, and 6.3. Only data for $\phi<60$\% are shown: we do
not study foams.

\begin{figure}[btp]\begin{center}
(a)\includegraphics[width=7.4cm]{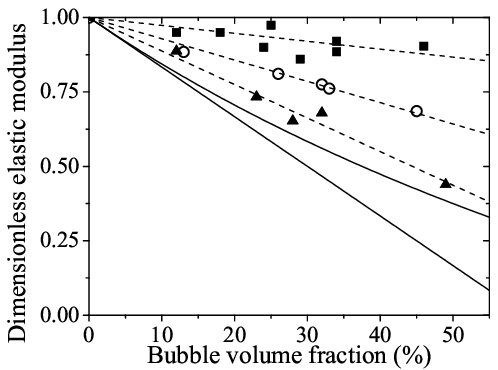}
(b)\includegraphics[width=7.4cm]{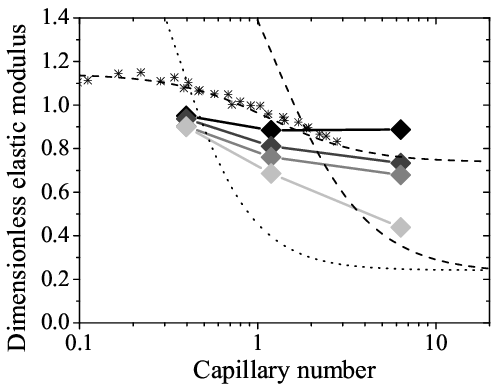} \caption{(a)
Dimensionless elastic modulus $g(\phi)=G'(\phi)/G'(0)$ vs. bubble
volume fraction $\phi$, for suspensions of bubble diameter
$d=$~100$\mu$m (squares), 300$\mu$m (empty circles), and 1.6~mm
(triangles), in an emulsion of elastic modulus $G'(0)=285$~Pa. The
full lines correspond to, from bottom to top: the theoretical
value of $g(\phi)$ for suspensions of fully deformable bubbles in
the dilute limit (Eq.~\ref{eq_linear_deformable_dilute}) and the
theoretical upper bound for isotropic suspensions of fully
deformable bubbles (Eq.~\ref{eq_linear_deformable_Mori}); the
dotted lines are linear fits to the data (their slopes are, from
top to bottom: -0.17, -0.72, -1.1). (b) Dimensionless elastic
modulus $g(\phi)$ vs. elastic capillary number $\Ca_{_G}$ at
different volume fraction (diamonds): $\phi\simeq11-12\%$ (black),
$\phi\simeq23-26\%$ (dark grey), $\phi\simeq32-34\%$ (medium
grey), $\phi\simeq45-49\%$ (light grey); crosses are dimensionless
viscosities measured in suspensions of bubbles in a Newtonian
fluid vs. viscous capillary number $\Ca_\eta$ at $\phi=11.5\%$
(data replotted from \citet{Rust2002a}); the dashed lines are the
phenomenological expression proposed by \citet{Rust2002a} (Eq.~9
of their paper) for $\phi=11.5$\% and $\phi=47.5$\%; the dotted
line is the phenomenological expression proposed by
\citet{Pal2004} (Eqs. 12 and~35-38 of the paper) for
$\phi=47.5$\%.}\label{fig_gprime_ttesbulles_emulconstante}
\end{center}\end{figure}

In Fig.~\ref{fig_gprime_ttesbulles_emulconstante}a, we observe
that the behavior depends significantly on $\Ca_{_G}$. In the
three studied systems, $g(\phi)$ decreases with $\phi$, and
decreases more and more with $\phi$ as $\Ca_{_G}$ is increased\footnote{Since the elastic moduli of the three systems are measured in the linear regime where their behavior does not depend on the strain amplitude, we also note that different $g(\phi)$ values can be obtained for a same value of $\Ca_\tau$ (see Eq.~\ref{eq_capillary_g_gamma}); this shows that $Ca_{\tau}$ is not a relevant parameter in this regime.}. A
linear fit to the data (with $g(0)=1$) yields slopes equal to -0.17, -0.72, and
-1.1 for increasing values of $\Ca_{_G}$; we recall that, for
fully deformable bubbles, the slope is expected to be equal to
-1.66 in the dilute limit.

The behavior at the lowest value of $\Ca_{_G}$ investigated here
(0.4) is close to a $g(\phi)\simeq1$ curve. This confirms the
observations on Fig.\ref{fig_gprime_ttesbulles_emulvariable},
where $g(\phi)\simeq1$ was obtained for $\Ca_{_G}\simeq0.5$ at
different bubble diameters (see Fig.~\ref{fig_capillary}). In this
case, the bubbles then seem to behave as equivalent elastic
spheres of elastic modulus equal to that of the interstitial
material; this may mean that in general bubbles behave as
equivalent elastic particles of modulus $G'_{eq}$ equal to the
capillary pressure $4\sigma_t/d$. At the highest value of
$\Ca_{_G}$ investigated, the $g(\phi)$ curve falls slightly above
the theoretical upper bound for bubbles with no surface tension
(Eq.~\ref{eq_linear_deformable_Mori}). This suggests that the
bubbles do not behave yet as fully deformable bubbles, but they
may be close to this asymptotic limit.

In Fig.~\ref{fig_gprime_ttesbulles_emulconstante}b, we now plot
the dimensionless elastic modulus $g(\phi)$ as a function of the
capillary number $\Ca_{_G}$ for different constant bubble volume
fractions $\phi$. Given the limited amount of data, and the
difficulty to target the exact same value of $\phi$ in different
systems, this plot is rather qualitative; data are indeed plotted
for $\phi$ values in four different intervals: between 11 and
12\%, between 23 and 26\%, between 32 and 34\%, and between 45 and
49\%. It is observed that $g(\phi)$ varies smoothly with
$\Ca_{_G}$. This evolution can be compared to that observed by
\citet{Rust2002a} for suspensions of bubbles in Newtonian fluids:
the dimensionless viscosities they have measured at $\phi=11.5\%$
are plotted vs $\Ca_\eta$ in
Fig.~\ref{fig_gprime_ttesbulles_emulconstante}b, together with the
empirical function they have proposed to account for their data
and for other data of the literature; the equation proposed by
\citet{Pal2004} is also shown. The \citet{Rust2002a} data decrease
more rapidly with $\Ca_\eta$ than ours do with $\Ca_{_G}$ for the
same value of $\phi$, and the $g(\phi)$ evolution predicted by
both the \citet{Rust2002a} and the \citet{Pal2004} empirical
equations at high $\phi$ is much more abrupt than that we observe.
Note however that several issues prevent from an in-depth
comparison. (i) \citet{Rust2002a} data were obtained on
polydisperse systems, and only up to $\phi=16\%$. (ii) We are not
aware of other data obtained at intermediate $\Ca_{_G}$ values in
the literature; at high $\phi$, the \citet{Rust2002a} and
\citet{Pal2004} equations have been fitted only to the asymptotic
viscosities corresponding to nondeformable and fully deformable
bubbles. It thus seems that the behavior of concentrated
suspensions of bubbles in a Newtonian fluid at capillary number of
order unity is not yet known. (iii) The two studied problems are
not exactly the same: in steady-state flows, bubbles are deformed
by shear and reach a steady shape. Here, the bubbles are only
slightly deformed by the low amplitude oscillatory measurement:
although they behave as soft deformable particles, they remain
basically spherical. Further theoretical work is needed to better
understand this difference.

The smooth variation of the dimensionless modulus with $\Ca_{_G}$
observed in Fig.~\ref{fig_gprime_ttesbulles_emulconstante}b
explains the apparent data collapse in the first series of
experiments (Fig.~\ref{fig_gprime_ttesbulles_emulvariable}):
although data corresponded to different capillary numbers
(Fig.~\ref{fig_capillary}), those were ranging only between 1 and
2.5; consistently, the -0.75 slope observed for the line onto
which data collapse in
Fig.~\ref{fig_gprime_ttesbulles_emulconstante}b is close to the
slope here observed for a constant $\Ca_{_G}$=1.2.

Note that it was not possible to study systems of lower $\Ca_{_G}$
value. Lower values of the bubble size would indeed pose finite
size effects problems as scale separation between bubbles/droplets
would not be ensured anymore. Emulsions of lower elastic modulus
should have a higher droplet size, which poses again finite size
effect problems, or a lower droplet volume fraction, which poses
the problem of reproducibility of the systems preparation (the
emulsion elastic modulus is  highly sensitive to a change of the
droplet volume fraction at the approach of its jamming packing
fraction). In future works, a possible way to prepare materials at
lower $\Ca_{_G}$ may be to use surfactants that would lower the
oil/water interfacial tension significantly more than the
air/water surface tension. It was also not possible to study systems of
higher $\Ca_{_G}$ value because bubbles tends to be broken
by the mixing process in such systems (see next section). In the
future, it will be necessary to use a new preparation method to
disperse homogeneously bubbles in an emulsion without mixing.

\section{Yield stress}\label{section_yield}

\subsection{Mixtures of foam and emulsion}\label{subsection_yield_mixes}

\begin{figure*}[t]\begin{center}
\hspace{0.5cm}(a)\hspace{-0.5cm}\includegraphics[width=12.8cm]{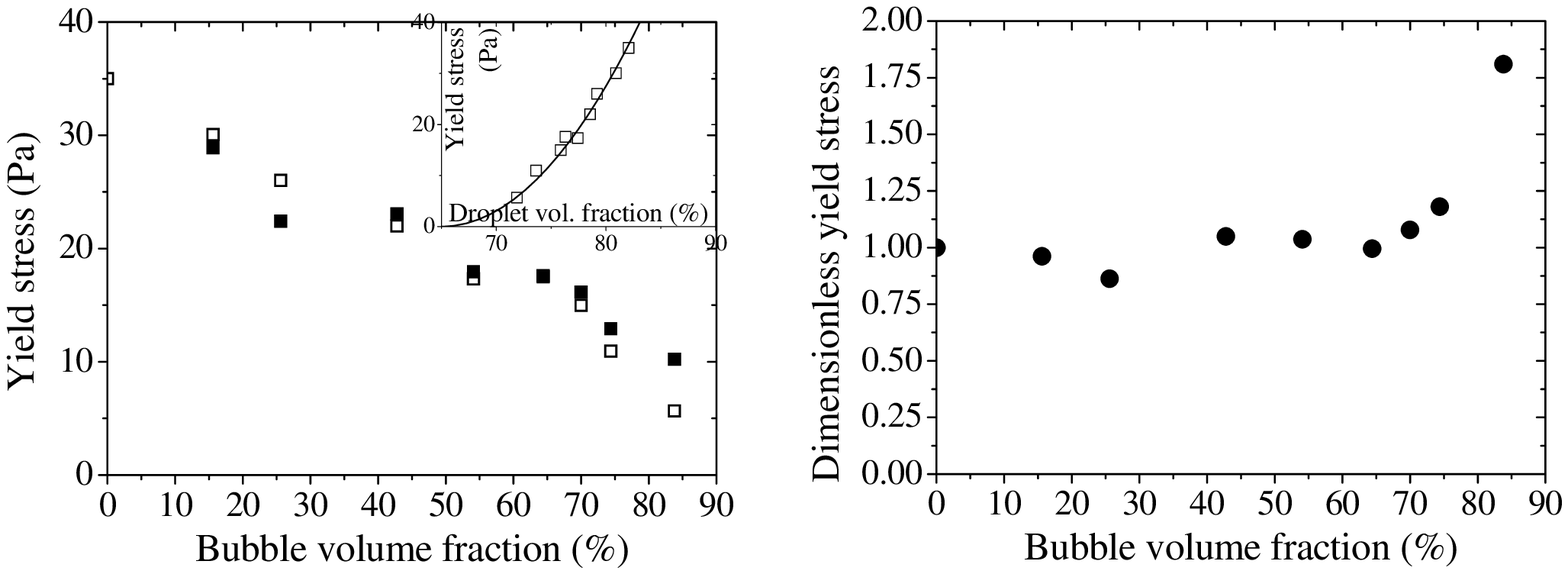}\hspace{-0.3cm}(b)\hspace{0.3cm}\, \\
\caption{(a) Yield stress vs. bubble volume fraction for the
suspensions of bubbles in the emulsion (filled squares), and for
the emulsions surrounding the bubbles in the suspensions (diluted
emulsions, empty squares).  (b) Dimensionless yield stress
$\tau_y(\phi)/\tau_{y,i}$ vs. bubble volume fraction $\phi$. The
bubble size is 320~$\mu$m. It is recalled that increasing $\phi$
corresponds to an increasing quantity of foam added to the
emulsion, and to a lower droplet volume fraction in the
interstitial emulsion. Inset: yield stress $\tau_{y,i}$ of the
interstitial emulsion vs. oil droplet volume fraction $\phi_d$ in
the emulsion. The line is a fit to the $(\phi_d-\phi_m)^2$ scaling
proposed by \citet{Mason1996} with
$\phi_m=65\%$.}\label{fig_tauy_phi_bulles320}
\end{center}\end{figure*}

As in Sec.~\ref{section_elastic}, we first study the cases in
which suspensions of increasing bubble volume fraction are
obtained by adding foam to a given emulsion in increasing foam to
emulsion mass ratios.

Fig.~\ref{fig_tauy_phi_bulles320}a shows the yield stress values
$\tau_y$ measured in suspensions of 320~$\mu$m bubbles and in
their corresponding interstitial emulsion, as a function of the
bubble volume fraction $\phi$ in the suspension. Because of its
dilution by the surfactant solution brought in by the foam, as
already observed for the elastic modulus, the yield stress of the
interstitial fluid decreases when the quantity of foam
incorporated in the emulsion is increased (i.e., when $\phi$
increases). The inset of Fig.~\ref{fig_tauy_phi_bulles320}a shows
the yield stress of the interstitial emulsion replotted as a
function of the droplet volume fraction in the emulsion; our data
are consistent with the observations of \citet{Mason1996} and with
the empirical law they propose to model the yield stress of
concentrated emulsions.

The yield stress of the bubble suspensions is found to decrease
with $\phi$ similarly to that of the interstitial emulsion. This
would mean that, as regards its plastic properties, the main
impact of adding a foam to the material is to dilute the
interstitial fluid. The impact of the presence of bubbles can be
more precisely evaluated by plotting the suspension dimensionless
yield stress $\tau_y(\phi)/\tau_{y,i}$ vs. bubble volume fraction
$\phi$ (Fig.~\ref{fig_tauy_phi_bulles320}b); here, $\tau_{y,i}$ is
the yield stress of the interstitial emulsion corresponding to
each suspension, i.e., accounting for dilution effects. In this
system (suspension of 320~$\mu$m bubbles), up to $\phi=80\%$,
$\tau_y(\phi)/\tau_{y,i}$ is found to remain basically constant
and equal to 1.  A significant increase of
$\tau_y(\phi)/\tau_{y,i}$ is observed only at the highest volume
fraction (85\%).

\begin{figure}[btp]\begin{center}
\includegraphics[width=7cm]{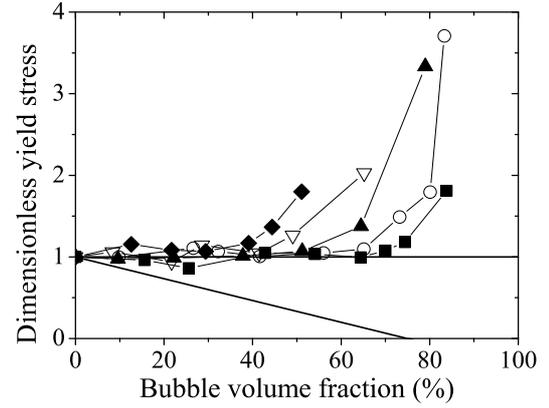}
\caption{Dimensionless yield stress $\tau_y(\phi)/\tau_{y,i}$ vs.
bubble volume fraction $\phi$, for suspensions of bubble diameter
$d=$~320$\mu$m (squares), 260$\mu$m (empty circles), 230$\mu$m (up
triangles), 210$\mu$m (empty down triangles), and 110$\mu$m
(diamonds). The full lines correspond to, from bottom to top: the
theoretical value of $\tau_y(\phi)/\tau_{y,i}$ for suspensions of
fully deformable bubbles in the dilute limit
(Eq.~\ref{eq_plastic_deformable_dilute}), and the theoretical
value of $\tau_y(\phi)/\tau_{y,i}$  for suspensions of
nondeformable bubbles in the dilute limit
(Eq.~\ref{eq_plastic_nondeformable_dilute}).}\label{fig_tauy_ttesbulles_emulvariable}
\end{center}\end{figure}

In Fig.~\ref{fig_tauy_ttesbulles_emulvariable}, we plot the
dimensionless yield stresses determined for all our suspensions,
of bubble diameter $d$ varying between 110 and 320~$\mu$m. Two
behaviors are observed. (i) For volume fractions $\phi$ lower than
a critical value $\phi'_c(d)$ that depends on $d$, all the data
seem to collapse onto a single line: $\tau_y(\phi)/\tau_{y,i}$ is
basically constant and equal to 1, i.e., the yield stress of the
suspensions is equal to that of their interstitial emulsion. (ii)
Above $\phi'_c(d)$, $\tau_y(\phi)/\tau_{y,i}$ is found to increase
with $\phi$, in a way that depends on $d$. In this last regime,
the yield stress of the suspension can be up to 3.7 times higher
than that of the interstitial emulsion. In the following, we first
focus on the first regime.

\subsection{Role of deformability}\label{subsection_yield_fixed_capillary}

As shown in Sec~\ref{section_theory}, a value of
$\tau_y(\phi)/\tau_y(0)\approx1$ is expected for nondeformable
bubbles in a yield stress fluid
(Eq.~\ref{eq_plastic_nondeformable_dilute}), whereas a strong
decrease (with a slope -4/3) should be observed for deformable
bubbles\footnote{We remind that this should be strictly true in
the dilute limit only: we do not yet have a prediction for all
values of $\phi$.} (Eq.~\ref{eq_plastic_deformable_dilute}).
Although the elastic modulus of these suspensions decreases with
$\phi$, this analysis suggests that, for $\phi<\phi'_c(d)$,
bubbles are not deformed at yield. This is fully consistent with
the values of the yield capillary number
$\Ca_{\tau_y}=\tau_y/(2\sigma_t/d)$ of the studied materials,
which are all small -- of order 0.01 to 0.1. It thus seems that
the elastic and plastic behaviors of the bubble suspensions are
governed by two independent capillary numbers.

\begin{figure}[btp]\begin{center}
\includegraphics[width=7cm]{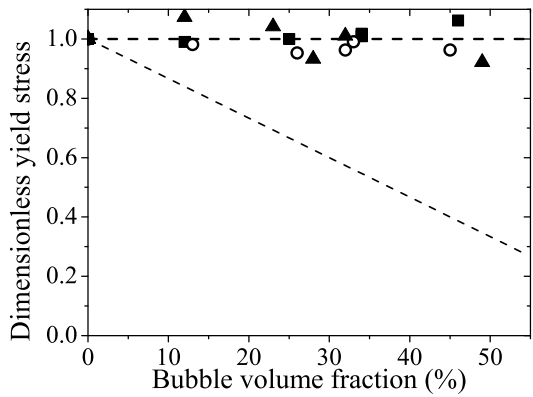}
\caption{Dimensionless yield stress $\tau_y(\phi)/\tau_y(0)$ vs.
bubble volume fraction $\phi$, for suspensions of bubble diameter
$d=$~100$\mu$m (squares), 300$\mu$m (empty circles), and 1.6~mm
(triangles), in an emulsion of yield stress $\tau_y(0)=9$~Pa. The
dashed lines correspond to, from bottom to top: the theoretical
value of $\tau_y(\phi)/\tau_y(0)$ for suspensions of fully
deformable bubbles in the dilute limit
(Eq.~\ref{eq_plastic_deformable_dilute}), and the theoretical
value of $\tau_y(\phi)/\tau_y(0)$  for suspensions of
nondeformable bubbles in the dilute limit
(Eq.~\ref{eq_plastic_nondeformable_dilute}).
}\label{fig_tauy_ttesbulles_emulconstante}
\end{center}\end{figure}

To confirm that the behavior is controlled by $\Ca_{\tau_y}$, we
now study suspensions of bubbles prepared with a same interstitial
emulsion surrounding the bubbles at any $\phi$ (see
Sec.~\ref{section_materials}). In
Fig.~\ref{fig_tauy_ttesbulles_emulconstante}, we first present the
dimensionless yield stresses measured in suspensions of 100$\mu$m,
300$\mu$m and 1.6~mm bubbles in a 9~Pa yield stress emulsion; the
corresponding plastic capillary numbers are 0.01, 0.04, and 0.2.
We observe that all data are consistent with
$\tau_y(\phi)/\tau_y(0)=1$, meaning that up to
$\Ca_{\tau_y}\simeq0.2$, bubbles are not deformed by shear.

\begin{figure*}[btp]\begin{center}
\includegraphics[width=16cm]{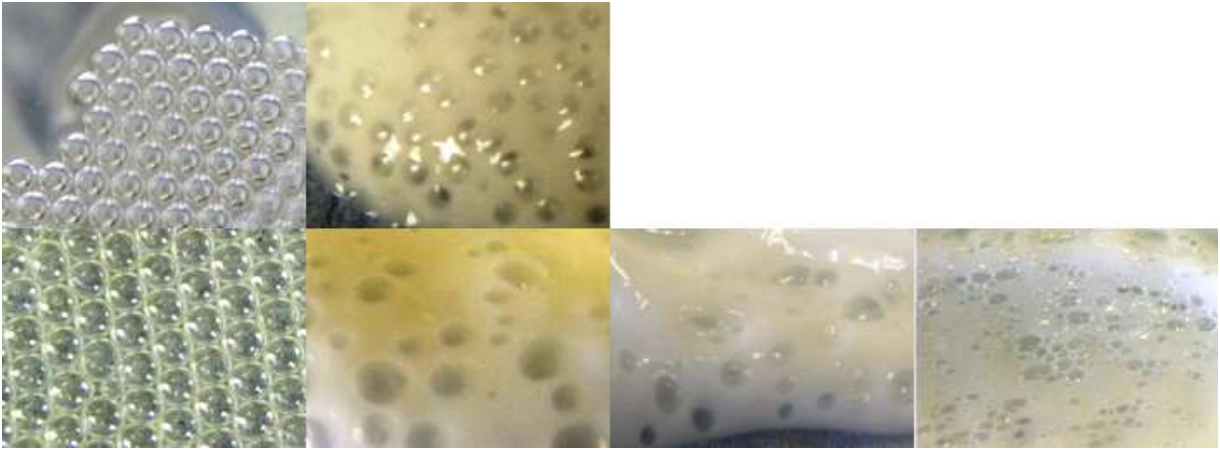}
\caption{Top: Pictures of an aqueous foam made of 1.6~mm diameter
bubbles (left) and of a suspension of 1.6~mm bubbles made by
mixing the foam with a 19~Pa yield stress emulsion (final
interstitial emulsion yield stress: 9~Pa) [same scale for both
images]. Bottom: Pictures of an aqueous foam made of 3~mm diameter
bubbles (left) and of a suspension of 3~mm bubbles made by mixing
the foam with (from left to right): a 19~Pa yield stress emulsion
(final interstitial emulsion yield stress: 9~Pa), a 87~Pa yield
stress emulsion (final interstitial emulsion yield stress: 19~Pa),
and a 87~Pa yield stress emulsion (final interstitial emulsion
yield stress: 72~Pa) [same scale for the 4
images].}\label{fig_photos_mixed_suspensions}
\end{center}\end{figure*}

To go one step further and to investigate the possibility of a
regime where bubbles are deformed by shear, we need to increase
the value of $\Ca_{\tau_y}$ as much as possible, which can be
obtained by increasing the material yield stress and the bubble
diameter. We have prepared materials with the goal of studying
suspensions of 3~mm bubbles in emulsions of 9, 19, and 72~Pa yield
stress (corresponding to target plastic capillary numbers of order
0.4, 0.8, and 3). In this end, foams of 3~mm bubbles were mixed
with emulsions of initial yield stress (before mixing) equal to 19
and 87~Pa (corresponding to initial capillary numbers of order 0.8
and 3.6). Photographs of the aqueous foam and of the prepared
bubble suspensions are shown in
Fig.~\ref{fig_photos_mixed_suspensions}; photograph of a
suspension of 1.6~mm bubbles is also presented for comparison.

In the case of the suspension prepared with an aqueous foam of
1.6~mm diameter bubbles, it is observed that the bubbles in the
suspension have the same size as in the aqueous foam. By contrast,
all suspensions prepared with an aqueous foam of 3~mm diameter
bubbles are polydisperse: small bubbles have appeared, due to
shear-induced bubble rupture during the mixing process (we checked
that the air content in the suspension matches that brought in by
the foam into the emulsion). Moreover, the bubbles are smaller for
higher values of the initial emulsion yield stress and of the
final interstitial emulsion yield stress. This is consistent with
a mechanism of shear-induced bubble breakup, which is expected to
depend on the ability of the bubbles to be deformed by shear. As
for bubble and droplet breakup in Newtonian fluids
\citep{Larson1999}, we expect this property to be governed by a
capillary number, the relevant one being here the plastic
capillary number $\Ca_{\tau_y}$\footnote{If mixing is rapid and if
viscous effects are important, for a constitutive behavior of the
form $\tau=\tau_y+f(\gdot)$, the relevant number might rather be
$\frac{\tau_y+f(\gdot)}{2\sigma_t/d}$.}. Consistently, in the
experiments of Fig.~\ref{fig_photos_mixed_suspensions}, the
initial values of the plastic capillary number leading to breakup
are of order 0.8 to 3.6. It is worth noting that the final
capillary numbers of the prepared suspensions, based on the
smallest bubble size observable on the pictures, is of order 0.2
in the 3 systems. Note also, that the final dimensionless yield
stress of the first 2 systems is of order 1 (we could not get a
reliable measurement of this property in the third system, which
was inhomogeneous after mixing). It thus appears that (i) for
$\Ca_{\tau_y}$ values of order 0.2 or less, the bubbles behave as
nondeformable inclusions under shear and that the suspension yield
stress is equal to that of the interstitial emulsion, and that
(ii) for initial values of $\Ca_{\tau_y}$ above a critical plastic
capillary number of order 0.2, bubbles are deformed and broken in
the sheared suspension. This might imply that, in any bubble
suspension prepared by mixing a foam and a paste, the final bubble
size in the system is always such that $\Ca_{\tau_y}\lesssim0.2$,
and that its yield stress is equal to that of the interstitial
paste. This possible important feature has to be studied in more
detail with well-controlled mixing procedures.

\subsection{Behavior at high $\phi$: `foam' regime and role of finite
size effects}

We finally come back onto the abrupt increase of the bubble
suspension yield stress observed at high volume fraction
(Fig.~\ref{fig_gprime_ttesbulles_emulvariable}). An increase of
$\tau_y$ with $\phi$ above a $d$-dependent critical value
$\phi'_c(d)$ is \textit{a priori} surprising as all our results
are consistent with bubbles behaving as rigid inclusions at low
and moderate $\phi$ values, the plastic capillary number being
small; the decrease of $\Ca_{\tau_y}$ with $\phi$ due to the
interstitial emulsion dilution is thus not expected to play any
role here, in contrast with what was observed with the elastic
modulus. Another mechanism should thus be proposed.

At high volume fraction, bubbles have to be deformed for geometric
reasons as in aqueous foams. This regime should typically occur
for $\phi>\phi_f$ with $\phi_f\approx64\%$ if films are allowed to
be as thin as possible. An important characteristic of the studied
system is that film thickness is here limited by the
microstructure of the emulsion: film thinning stops when the oil
droplets cannot be expelled anymore, which occurs when the film
thickness is of the order of a few droplet diameters
\citep{Goyon2010}. It is expected that flow of the confined
emulsion is harder to enforce than that of the bulk material, due
to its lack of disorder; this would explain the increase in the
suspension yield stress when reaching this `foam' regime. By
contrast, this should not affect the elastic properties of the
material. As the minimum distance between bubbles is fixed by the
oil droplet size, the maximum packing fraction $\phi_f(d)$
delimiting the `suspension' and `foam' regimes should depend on
the bubble diameter $d$. Noting $e$ the minimum film thickness and
$d_d$ the oil droplet diameter, and assuming $\phi_f\approx64\%$
when $d>>d_d$, $\phi_f(d)$ can actually be evaluated as
$\phi_f(d)=0.64/(1+e/d)^3$. Assuming a minimum film thickness
$e\approx4d_d$ \citep{Goyon2010}, this yields $\phi_f$ values
ranging between 42\% for the 110~$\mu$m bubbles and 55\% for the
320~$\mu$m bubbles. Note that the proposed mechanism is probably
not specific to suspensions of bubbles in emulsions: the same
might occur with any paste since their microstructure is often of
the order of 1~$\mu$m.

This minimum allowable film thickness has another important
consequence, as pointed out by \citet{Goyon2010}: bubble
suspensions cannot exist at volume fractions higher than a
$d$-dependent value $\phi_m(d)$ that is fixed by the minimum film
thickness $e$. This is the `dry' limit of `foamy yield stress
fluids'. Assuming a structure of Kelvin cells in this limit,
\citet{Goyon2010} have evaluated $\phi_m\approx1-1/(1+0.3d/e)$;
this would yield $\phi_m$ values ranging between 66\% for the
110~$\mu$m bubbles and 85\% for the 320~$\mu$m bubbles. This might
explain why could not explore the same ranges of volume fractions
for different bubble sizes in our experiments (see
Figs.~\ref{fig_gprime_ttesbulles_emulvariable}
and~\ref{fig_tauy_ttesbulles_emulvariable}): we did not manage to
study suspensions of 110~$\mu$m bubbles at volume fractions higher
than $\simeq$ 55\%, whereas suspensions of 320~$\mu$m bubbles were
studied up to $\simeq$~85\%.

\section{Conclusion}\label{section_conclusion}

We have studied the rheological behavior of mixtures of foams and
pastes. We have designed well-defined systems by mixing
monodisperse foams and monodisperse concentrated emulsions,
characterized by large bubble to oil droplet size ratios. These
materials are model suspensions of bubbles in a yield stress
fluid. We have shown that the elastic and plastic behaviors of
these materials are governed by two different dimensionless
numbers: the elastic capillary number $\Ca_{_G}$, which is the
ratio of the paste elastic modulus to the bubble capillary
pressure, and the plastic capillary number $\Ca_{\tau_y}$, which
is the ratio of the paste yield stress to the bubble capillary
pressure.

In our systems, $\Ca_{_G}$ ranges from 0.3 to 10 and the
dimensionless elastic modulus of the material decreases with the
bubble volume fraction: bubbles behave as soft elastic inclusions.
This decrease is all the sharper as $\Ca_{_G}$ is high, which
accounts for the softening of the bubbles as compared to the
paste. The transition from soft to stiff bubbles seems to occur at
$\Ca_{_G}\simeq0.4$. When mixing a foam and a paste, in some
conditions, two contradictory effects might thus be observed as
$\phi$ is increased: (i) a decrease of the elastic modulus due to
the dilution of the interstitial material by the surfactant
solution brought in by the foam, and (ii) an increase of the
dimensionless elastic modulus due to the consequent decrease of
$\Ca_{_G}$, if values lower than $\simeq0.4$ are reached. Further
investigation is needed to cover a wider range of $\Ca_{_G}$
values and in particular to characterize the regime of stiff
bubbles. Theoretical developments based on micromechanical
approaches are in progress and should allow the modelling of the
elastic modulus as a function of the bubble volume fraction and of
the elastic capillary number.

For $\Ca_{\tau_y}$ values lower than $\simeq0.2$, the
dimensionless yield stress of the suspensions is found to be
constant and equal to 1 in most cases, consistent with our
predictions for the nonlinear behavior of suspensions of
nondeformable bubbles. We have tried to prepare systems with a
target value of $\Ca_{\tau_y}$ larger than 0.2. Bubble breakup is
observed during mixing of these systems, and the final bubble size
seems to be set by the paste yield stress: it is smaller when the
paste yield stress is higher. This preliminary result is of high
practical importance: it might imply that, in any bubble
suspensions prepared by mixing a foam and a paste, the final
bubble size in the system is always such that
$\Ca_{\tau_y}\lesssim0.2$, and that its yield stress is equal to
that of the interstitial paste. Shear-induced bubble breakup has
to be studied in more detail as a function of shear history, to
better understand which suspensions of bubbles are produced by a
given mixing process. To perform such studies, new procedures are
needed to disperse homogeneously unbroken bubbles of controlled
size in a given yield stress fluid at high values of the plastic
capillary number.

At high bubble volume fraction $\phi$, we have observed a regime
where the yield stress increases abruptly with $\phi$, the
transition volume fraction being lower for lower bubble to droplet
size ratio. This is understood as a `foamy yield stress fluid'
regime, where the paste mesoscopic constitutive elements (here,
the oil droplets) are strongly confined in the films between the
bubbles. We are currently conducting further investigations to
better characterize and understand this regime, which is of the
highest importance for practical issues.

\begin{acknowledgements}
We thank Mohammed Bouricha for help on some of the experiments.\\
We acknowledge funding from Saint-Gobain Recherche.
\end{acknowledgements}

\end{document}